\documentclass[12pt]{article}
\usepackage{amsmath}
\usepackage{amsthm}
\usepackage{amsfonts}
\usepackage{amssymb}
\usepackage{amstext}
\usepackage{amsopn}
\usepackage{graphicx}
\usepackage{subfig}
\input epsf

\newtheorem{thm}{Theorem}

\newtheorem{lemma}{Lemma}

\newtheorem{preremark}{Remark}  
\newenvironment{remark}{\begin{preremark}\rm}{\end{preremark}}

\begin{document}
 \def\today{23~November~2011, revised 14~January~2012}
 \def\Question#1{\textbf{Question}: #1}
 \def\Answer  #1{\hfill\break\textbf{Answer:} #1\medskip}

\title{\bf 
Exact meromorphic stationary
solutions of the real cubic Swift-Hohenberg equation}
\author{\textsc{Robert Conte${}^{1,2}$\thanks{
Partially supported by PROCORE - France/Hong Kong joint research grants F-HK29/05T, F-HK39/11T
and 
RGC grant HKU 703807P. 
\mbox{\hspace{4truecm}}} 
, Tuen-Wai Ng${}^{1}$\thanks{
Partially supported by PROCORE - France/Hong Kong joint research grants F-HK29/05T, F-HK39/11T
 and 
RGC grant HKU 703807P. 
\mbox{\hspace{4truecm}}} 
and Kwok-Kin Wong${}^{1}$\thanks{
Partially supported by RGC grant HKU 703807P and a post-graduate studentship at HKU.
\mbox{\hspace{0.9truecm}}}}}
 
\date{}
\maketitle

\centerline{\textbf{\today}}

\begin{figure}[b]
\rule[-2.5truemm]{5cm}{0.1truemm}\\[2mm]
{\footnotesize  
2000{\it Mathematics Subject Classification: Primary} 35Q53.
\par\noindent 1.
Department of Mathematics,
The University of Hong Kong,
Pokfulam Road.
\smallskip
\par\noindent 2.
LRC MESO,
Centre de math\'ematiques et de leurs applications (UMR 8536) et CEA-DAM,
\\ \'Ecole normale sup\'erieure de Cachan, 61, avenue du Pr\'esident Wilson,
\\ F--94235 Cachan Cedex, France.
\smallskip
\par\noindent E-mail: Robert.Conte@cea.fr, ntw@maths.hku.hk, wkkm@graduate.hku.hk
\par
}

\end{figure}

\begin{quotation}
\noindent{\textbf{Abstract}.  
We show that all meromorphic solutions of the stationary reduction
of the real cubic Swift-Hohenberg equation are elliptic or degenerate elliptic. 
We then obtain them all explicitly
by the subequation method,
and one of them appears to be a new elliptic solution.
}
\end{quotation}

\def \D {\hbox{d}}
\def\CRAS{C.~R.~Acad.~Sci.~Paris}


\section{Introduction}
\label{secionIntro}

The real cubic Swift-Hohenberg (RCSH) equation 
\begin{equation}
\frac{\partial u}{\partial t}=\varepsilon u
-\left(1+\frac{\partial^2}{\partial x^2}+\frac{\partial^2}{\partial y^2}
 \right)^2 u
-u^3, 
\qquad \varepsilon \in \mathbb{R}
\label{PDE_RCSH0}
\end{equation} 
originally proposed in \cite{SH1977}, is a model for 
Rayleigh-B\'enard convection in hydrodynamics.
Since then, the equation and its generalizations have also been used 
in various areas,
in particular those where the field $u$ needs to be a complex amplitude,
such as laser \cite{LMN} and nonlinear
optics \cite{LG1996}. 
Much attention of the researches about RCSH has been put into its relation 
to the theory of pattern formation. 
The phenomena of pattern formation occur in a large variety of physical 
systems. 
We refer the readers to the extensive review \cite{CrossHohenberg1993} 
and the monographs \cite{CG2009} and \cite{Hoyle2006}.

The Swift-Hohenberg equation (\ref{PDE_RCSH0}) admits a 
stationary reduction by taking $u(x,y)=\sqrt{\varepsilon-1}U(Z)$,
with the rescaling $Z=(\varepsilon-1)^{1/4} (x+y)/2$:
\begin{equation}
U''''+a U''+ U^3-U=0,\ ':=\frac{\D}{\D Z}.
\label{ODE_RCSH0}
\end{equation}
Here $a=2/\sqrt{\varepsilon-1}$. 
The ODE (\ref{ODE_RCSH0})
is sometimes mentioned as
``extended Fisher-Kolmogorov equation'' \cite{vandenBerg1998,Kwapisz2000}
for $a<0$.

The equation (\ref{ODE_RCSH0}) has attracted intensive studies concerning, 
for example, the existence of various types 
of solutions and qualitative properties of the solution curves 
\cite{vdBPT2001,vandenBerg2003,Hoyle2006,Kwapisz2000,PT2001,PR2004,SW2009}.
For instance, 
when $U$ is a smooth function defined on the real line such that 
$U-1 \in H^2({\mathbb R})$, 
Santra and Wei \cite{SW2009} proved the existence of a homoclinic solution $U$
for each $a \in[0,a_0]$, where $a_0=1.228\dots$.\footnote{
The value $a_0$ here is different from that in \cite{SW2009}.
In \cite[p.~2040]{SW2009}, $a_0=\beta_0^2$ where $\beta_0=\sqrt{\sqrt{2}/k_0}$,
and $4k_0^2-2k_0-3=0$. 
The value $a_0$ stated in \cite{SW2009} is probably a typographical error.} 
\label{pagefootnote} 

We shall see in section \ref{sectionComparison} 
that a meromorphic homoclinic solution can only exist when 
$a=10/\sqrt{11}=3.015\dots$, 
which is outside the above specified range,
and we shall write down this exact meromorphic
homoclinic solution in terms of the $\tanh$ function (i.e. solution (\ref{RSHsol2fam_trigoU})). 
Indeed, in this paper we shall find all the  meromorphic solutions 
of (\ref{ODE_RCSH0}) explicitly. 

The method to derive all meromorphic solutions of (\ref{ODE_RCSH0}) 
is twofold.
We first use an analysis based on Nevanlinna theory proposed by Eremenko 
\cite{EremenkoKS} and slightly modified in \cite{ConteNg2010} 
to prove that any meromorphic solution is necessarily elliptic
or degenerate of elliptic.
In a second step, 
we use the subequation method first proposed in \cite{MC2003} 
to characterize each elliptic or degenerate elliptic solution
by some first order ODE,
which are easily integrated by classical methods.

The overall advantage of the method is that, once the algorithm is completed, 
all possible meromorphic solutions are exhausted. 
The detailed procedure of the method can be found in \cite{ConteNg2010}.

In section \ref{sectionClassW},
we show that all meromorphic solutions of (\ref{ODE_RCSH0}) belong to class $W$
(like Weierstrass), 
which consists of elliptic functions and their successive degeneracies,
i.e.: 
elliptic functions, 
rational functions of one exponential $\exp(kz), k\in\mathbb{C}$
and rational functions of $z$. 
We then apply the subequation method to find explicitly all these class $W$
solutions in section \ref{sectionMeromorphic_solutions}. 
Finally, in section \ref{sectionComparison}, 
we shall compare the meromorphic solutions obtained 
in section \ref{sectionMeromorphic_solutions} 
with some known results concerning (\ref{ODE_RCSH0}).

Let us finally mention a recent paper \cite{ChiangHalburd} which uses
similar ideas.
Given a specific nonlinear second order ODE,
these authors use Wiman-Valiron theory combined with local series
analysis to explicitly determine all entire solutions.

\newpage
\section{Class $W$ solutions}
\label{sectionClassW}

Only five polynomial terms $(U U'', U''', U^2, U', 1)$
can be added to (\ref{ODE_RCSH0})
while retaining the double pole behaviour displayed below, Eq.~(\ref{SHlau}).
For a reason to be explained soon,
we only retain two of them $(U^2,1)$
and consider the 
more general form of the equation (\ref{ODE_RCSH0})
in which the three fixed points $U=-1,0,1$ are not necessarily equispaced, 
\begin{equation}
u''''+60 b u'' -\frac{120}{c^2} \left( u^3-3 s_1 u^2 +3s_2 u -s_3 \right)=0,\
':=\frac{\D}{\D z}. 
\label{ODE_RCSH4}
\end{equation}
The complex constant $c$, which could be scaled out,
will be useful for parity considerations, for example
to determine the second Laurent series (\ref{SHlau})
by simply changing $c$ to $-c$ in the first one. 

Its three fixed points $a_j$ are defined by
\begin{equation}
u^3-3 s_1 u^2 +3s_2 u -s_3 = (u-a_1)(u-a_2)(u-a_3)
\label{ODE_RCSH_aj}
\end{equation}
When the three fixed points $a_j$ of (\ref{ODE_RCSH4}) are equispaced,
\begin{equation}
3 s_1 s_2 - 2 s_1^3 - s_3=0,\
(2 a_1-a_2-a_3) (2 a_2-a_3-a_1) (2 a_3-a_1-a_2)=0,
\label{ODE_RCSH_equispaced}
\end{equation}
equation (\ref{ODE_RCSH4}) is indeed equivalent to (\ref{ODE_RCSH0}) 
under the rescaling, 
\begin{eqnarray}
& &
U=\lambda (u-s_1),\ z= \mu Z,\ 
\nonumber \\ & &
\lambda^2=-\frac{1}{3 (s_2-s_1^2)}=\frac{6}{(a_2-a_3)^2+(a_3-a_1)^2+(a_1-a_2)^2},\ 
\mu^2=\frac{a}{60 b}.
\label{scalingtransfo}
\end{eqnarray}

Equation (\ref{ODE_RCSH4}) admits the integrating factor $u'$, 
thus yielding the first integral
\begin{equation}
K = u' u'''-\frac{1}{2} {u''}^2+30 b {u'}^2
-\frac{30}{c^2} \left(u^4-4 s_1 u^3+6 s_2 u^2-4 s_3 u 
                     + 3 s_1^4  -6 s_1^2 s_2 +4 s_1 s_3 \right),
\label{ODE_RCSH3}
\end{equation}
where $K$ is the integration constant, 
and the added constant terms will make later expressions simpler. 

\remark
The reason for excluding terms $(u u'', u''', u')$ in (\ref{ODE_RCSH4})
is to allow the first integral (\ref{ODE_RCSH3}) to exist.
\end{preremark}

Let $u$ be a meromorphic solution (defined on ${\mathbb C}$) of (\ref{ODE_RCSH3}). 
To find all such $u$, we shall first prove the following result.

\begin{thm}
If the ODE (\ref{ODE_RCSH3}) has a particular meromorphic solution $u$, 
then $u$ belongs to the class $W$.
\label{Theorem1}
\end{thm}

The proof requires the use of the Nevanlinna theory,
whose main features required here are introduced below. 
Some good references of Nevanlinna theory are \cite{Hayman} and 
\cite{LaineBook}. 

The method for proving \ref{Theorem1} 
comes from a paper of Eremenko \cite{Eremenko1982} 
which shows that all meromorphic solutions of odd order Briot-Bouquet 
differential equations 
with at least one pole must belong to class $W$. 
The even order case was recently proven to be also true \cite{ELN}. 
Eremenko's method applied to (\ref{ODE_RCSH3}) or other autonomous nonlinear 
algebraic ODEs consists of two main steps:

\begin{enumerate}
\item
Show that there are {\it finitely many} 
Laurent series with a pole at any point $z_0$ which satisfy (\ref{ODE_RCSH3}).

\item
Show that any {\it transcendental} meromorphic solution $u$ of (\ref{ODE_RCSH3}) 
must have infinitely many poles.
\end{enumerate}

Step 1 usually involves computing the Fuchs indices of (\ref{ODE_RCSH3}) and 
Step 2 requires applying Nevanlinna theory or Wiman-Valiron theory. 
Once we have established the two steps, 
we can then show easily that the solution $u$ must be periodic 
and hence belong to class $W$.

For the convenience of readers, we shall include a proof of 
\ref{Theorem1} 
which is very similar to the one given in \cite{EremenkoKS} 
or \cite{ConteNg2010}.

We first introduce some notations commonly used in Nevanlinna theory
(see \cite{Hayman}, also see \cite{LaineBook} for a quick introduction). 
Assume $f$ to be a non-constant meromorphic function on 
the open disc $D(r)$ where $r$ can be $\infty$.

Denote the number of poles of $f$ on the closed disc 
$\overline{D}(r)$ by $n(r,f)$, counting multiplicity. 
The function $n(r,f)$ is usually called the \textbf{unintegrated counting function}.
Note that $n(r,\frac{1}{f})$ would be the number of 
zeros of $f$ on $\overline{D}(r)$ and $n(r,\frac{1}{f-a})$ 
would be the number of times $f$ takes $a$.

Define the \textbf{integrated counting function} $N(r,f)$ as
\begin{equation}
N(r,f)=n(0,f)\log r +\int^{r}_{0} \left[n(t,f)-n(0,f)\right] \frac{dt}{t},
\end{equation}
and the \textbf{proximity function} $m(r,f)$ as
\begin{equation}
m(r,f)=\int^{2\pi}_{0}\log^{+} f(re^{i\theta})\frac{d\theta}{2\pi},
\end{equation}
where $\log^{+}x=\max {\left\{0,\log x\right\}}$.

Finally, we define the Nevanlinna's characteristic function $T(r,f)$ as
\begin{equation}
T(r,f)=m(r,f)+N(r,f).
\end{equation}

A basic fact concerning the functions $T,N,m$ is the following:
\begin{thm}[Nevanlinna's first fundamental theorem] 
Let $f$ be a meromorphic function and $a\in \mathbb{C}$. Then
$$ T(r,\frac{1}{f-a})=m(r,f)+N(r,f)+O(1) $$
as $r\rightarrow \infty$.
\label{Nevan1}
\end{thm}

We also need the following well known results (detailed discussions
can be found in \cite{LaineBook}): 

\begin{lemma}
A meromorphic function $f$ is rational if and only if $T(r,f)=O(\log{r})$.
\label{lem1}
\end{lemma}

\begin{lemma}[Clunie's lemma] 
Let $f$ be a transcendental meromorphic solution of 
\begin{equation} f^n P(z,f,f',...)=Q(z,f,f',...),\end{equation} 
where 
$n$ is a nonzero positive integer, 
$P$ and $Q$ are polynomials in $f$ 
and its derivatives with meromorphic coefficients 
$\{a_\lambda|\lambda \in I\}$, such that 
for all $\lambda \in I$, $m(r,a_{\lambda})=S(r,f)$,
where $I$ is some known index set. 
If the total degree\footnote{
Defined as the global degree in all derivatives $f^{(j)}, j \ge 0$.}  
of $Q$ as a polynomial in $f$ 
and its derivatives is less than or equal to $n$, then 
\begin{equation}m(r,P(z,f,f',...))=S(r,f).\end{equation}
\label{lem2}
\end{lemma}
Here $S(r,f)$ is  called the ``small'' function and
$m(r,P(z,f,f',...))=S(r,f)$ means that the function on left hand
side has growth $o(T(r,f))$ as $r\rightarrow \infty$ outside a
possible exceptional set of finite linear measure. 

\medskip
\noindent
\textbf{Proof of Theorem \ref{Theorem1}}.
Let's begin with Step 1. 
Suppose $u$ is a meromorphic solution of (\ref{ODE_RCSH3}).
If $u$ has a movable 
pole at $z=z_0$, then it must be a double pole. Let
\begin{equation}
u(z)=\alpha_0(z-z_0)^{-2}+\alpha_1(z-z_0)^{-1}+\alpha_2+\dots 
\label{SHlau}
\end{equation}
be the Laurent series of $u$ in a neighborhood of $z_0$. 
Inserting 
(\ref{SHlau}) into (\ref{ODE_RCSH3}) and balancing the leading terms, 
we get $\alpha_0=\pm c$.
Using for example the procedure given in \cite{Cargese1996Conte}, 
the Fuchs indices of the ODE (\ref{ODE_RCSH3}) are found to be 
independent of the sign of $\alpha_0$ and equal to 
$-1$, $(7\pm \sqrt{-71})/2$. 
This implies that all other Laurent series coefficients in (\ref{SHlau}) 
are uniquely determined by the leading coefficients $\alpha_0$ 
and are independent of $z_0$ \cite{Cargese1996Conte},
thus we only have two distinct Laurent series. 
This shows that there exist at most two meromorphic functions 
with a pole at $z=z_0$ satisfying (\ref{ODE_RCSH3}).

If $u$ is rational, then $u$ belongs to $W$ and we are done. 
Now assume $u$ to be transcendental. 
By putting (\ref{ODE_RCSH3}) into the form:
\begin{equation}
\frac{30}{c^2}u^4=u' u'''-\frac{1}{2}u''^2+30 b u'^2
-\frac{30}{c^2} \left(-4 s_1 u^3+6 s_2 u^2-4 s_3 u \right)
+\hbox{constant},
\end{equation}
we conclude from Clunie's lemma ($f=u,n=3,P=u$) 
that $m(r,u)=S(r,u)$ and hence $(1-o(1))T(r,u)=N(r,u)$. 
If $u$ has only finitely many poles, 
then $N(r,u)=O(\log r)$ and hence $T(r,u)=O(\log r)$. 
By Lemma \ref{lem1}, 
$u$ must be rational, which is a contradiction. 
Therefore $u$ must have infinitely many poles and this completes Step 2.

We are now ready to prove that $u$ belongs to class $W$. 
We first claim that, if $u$ is transcendental, it must be periodic. 
Suppose $u$ has a pole at $z=z_0$. 
By the previous analysis, there exist at most two meromorphic solutions
of (\ref{ODE_RCSH3}) with poles at $z=z_0$. 
Let $z_j, j=1,2,...$, be the infinitely many poles of $u(z)$; 
then for all $j=1,2,\dots$, $w_j(z)\equiv u(z+z_j-z_0)$ are also
solutions of (\ref{ODE_RCSH3}) with a pole at $z_0$.

Since there exist at most two Laurent series around the pole $z_0$,
some of the $w_j$'s must be the same, otherwise we get a contradiction to
the maximum number of possible Laurent series.
But this implies that for some $j \neq i$, $u(z-z_0+z_j)\equiv u(z-z_0+z_i)$ 
and hence $u(z)\equiv u(z+z_i-z_j)$ in a neighborhood of $z_0$.
Since $u$ is meromorphic, we can conclude that $u$ is periodic in $\mathbb{C}$
with period $z_i-z_j$.

By a suitable rescaling, we may assume that $2\pi i$ is a primitive period 
of $u$. 
Let $D=\{z:0\leq \Im z<2\pi\}$.  
If $u$ has more than two poles in $D$, we can also consider the solutions
defined by $w_j(z)=u(z+z_j-z_0)$ similar to the last paragraph. 
Then since the number Laurent series is at most two, for some $j \neq i$, 
$u(z-z_0+z_j)\equiv u(z-z_0+z_i)$ and $u$ must be periodic in $D$ with 
period $z_i-z_j$. Note that $z_i-z_j\neq 2k\pi i$ for any $k\in \mathbb{Z}$
because $z_i,z_j \in D$. Thus $u$ is doubly periodic and therefore elliptic.

If $u$ has one or two poles in $D$,
then since $u$ is a periodic function with period $2\pi i$, 
we have $N(r,u)=O(r)$ as $r\rightarrow \infty$. 
Since $(1-o(1))T(r,u)=N(r,u)$, we have $T(r,u)=O(r)$. 
By the Nevanlinna's first fundamental theorem, 
for any $\zeta\in \mathbb{C}$, $N(r,1/(u-\zeta))=O(r)$ as $r\rightarrow \infty$.
Since $u$ is periodic with period $2\pi i$ and $N(r,1/(u-\zeta))=O(r)$, 
it must take each $\zeta$ only finitely many times in $D$. 
This implies that the function $R(z)=u(\ln z)$ is a single-valued 
analytic function 
defined in the punctured complex plane $\mathbb{C}-\{0\}$ 
which takes each $\zeta$ only finitely many times. 
So $0$ is a removable singularity of $R$. 
This implies that $R$ is rational. 
Since $u(z)=R(e^z)$, 
we conclude that $u$ must belong to the class $W$ and \ref{Theorem1}
is proven. 

\remark
From the proof,
we conclude that, if the solution $u$ is elliptic,
the number of poles of $u$ in the fundamental parallelogram must
be at most two.
\end{preremark}

\remark
The ODE (\ref{ODE_RCSH4}) admits infinitely many Laurent series
parametrized by the arbitrary constant $K$,
therefore \ref{Theorem1}
cannot apply to it.
\end{preremark}

\section{Explicit meromorphic solutions}
\label{sectionMeromorphic_solutions}

On the one hand, Theorem \ref{Theorem1} 
shows that any meromorphic solution of (\ref{ODE_RCSH3}) 
must be in the class $W$. 
On the other hand,
there exists a method (the subequation method proposed in \cite{MC2003})
which can be applied to find all those solutions of an algebraic ODE which belong to class $W$.
Now, combining these two features allows one to find in closed form
all the particular solutions of (\ref{ODE_RCSH3}) which are meromorphic.

Let us first recall the classical definition of the 
\textit{elliptic order} of an elliptic function: 
this is the common number of poles or zeros, counting multiplicity, 
inside a fundamental parallelogram.
The method is based on the following well known theorem of Briot and Bouquet 
on first order Briot-Bouquet differential equations \cite{BB,Hille}:

\begin{thm}
Any elliptic function obeys a first order algebraic differential equation 
of the form 
\begin{eqnarray}
F(u,u') \equiv
 \sum_{k=0}^{m} \sum_{j=0}^{2m-2k} a_{j,k} u^j {u'}^k=0,\ 
a_{0,m}\not=0,\ 
\label{1ODE}
\end{eqnarray}
where $m$ is the elliptic order of $u$.
\end{thm}
In the case of (\ref{ODE_RCSH3}), if the solution $u$ is elliptic, 
it has either one double pole 
(then $m=2$, and only one Laurent series contributes to this solution) 
or two double poles (then $m=4$, and the two Laurent series contribute to this solution)
in a fundamental parallelogram.
Therefore the subequation method must be applied with, successively, $m=2$ and $m=4$.

If $u$ is a degenerate elliptic function, 
the above theorem does not apply,
but any Laurent series expansion around a pole still has the polar order two 
and there exist at most two such expansions,
therefore $u$ must still obey an ODE of the form (\ref{1ODE}), with $m=2$ or $4$.
Now let $u$ be a meromorphic solution of (\ref{ODE_RCSH3}). 
{}From section \ref{sectionClassW}, 
we know that we can determine uniquely and recursively the coefficients 
of the two possible Laurent series of $u$ in a neighborhood 
of a movable pole $z=z_0$. 
After putting these Laurent series into (\ref{1ODE}), 
the equation (\ref{1ODE}) should vanish identically. 
This generates a countably infinite (hence overdetermined and easy to solve) 
system of linear equations in the finitely many unknowns $a_{j,k}$'s. 
By solving it for the $a_{j,k}$'s, 
the first order ODEs $F(u,u')=0$ are known explicitly 
and all admit (\ref{ODE_RCSH3}) as a differential consequence. 
This is why we call the obtained first order ODEs $F(u,u')=0$ 
a \textit{subequation} of (\ref{ODE_RCSH3}). 
For details of the implementation of the algorithm, see \cite{MC2003}.

\subsection{Subequations obeyed by one Laurent series} 

These subequations have degree $m=2$. 
For a given Laurent series (\ref{SHlau}), 
the method yields one subequation, 
at the price of one constraint among the fixed coefficients 
$K,b,c,s_1,s_2,s_3$ of (\ref{ODE_RCSH3}), 
\begin{eqnarray} 
& & {\hskip -6.0 truemm}
\left\lbrace 
\begin{array}{ll}
\displaystyle{
F_2 \equiv c {u'}^2 -4 (u-s_1 -b c)^3 +20 (s_1^2 -s_2 -b^2 c^2)(u-s_1-b c) 
}\\ \displaystyle{
\phantom{S_2:\ F \equiv} 
  + 10 (s_3 - 3 s_1 s_2  + s_1^3 + 2 b c s_2 - 2 b c s_1^2 + 4 b^3 c^3)=0, 
}\\ \displaystyle{
c^2 K=10 (-128 b^4 c^4 -88 b^2 c^2 (s_2-s_1^2) 
   + 18 b c (-2 s_1^3 +3 s_2 s_1- s_3)).
}
\end{array}
\right.
\label{RSHP_subeq_1fam}
\end{eqnarray}
The solution for the other Laurent series is obtained by changing 
$c$ to $-c$. 
This first order ODE (\ref{RSHP_subeq_1fam}) is nothing else than 
the canonical equation of Weierstrass, 
up to some translation and rescaling.

This defines the codimension-one\footnote{
Following a frequent terminology,
we define the \textit{codimension} of a solution of a
differential equation as the number of constraints required
among the parameters of the equation.
}
 solution of (\ref{ODE_RCSH3}) 
\begin{equation}
\begin{cases}
   u = s_1 + b c + c \wp(z-z_0,g_2,g_3),\ 
\\ g_2= 20 (s_1^2 - s_2 - b^2 c^2)/c^2,\
\\ g_3= 10 (s_3 -3 s_1 s_2 + s_1^3 + 2 b c s_2 - 2 b c s_1^2 + 4 b^3 c^3)/c^3.
\end{cases}
\label{RSHsol1fam_elliptic}
\end{equation}
This solution is not new, 
and has already been obtained \cite{MAA}
by assuming $u$ (which has movable double poles)
to be a polynomial in $(\wp,\wp')$ having double poles, 
i.e.~reducing to an affine function of $\wp$.

When the genus of the curve $F_2(u,u')=0$ is zero,
i.e.~when $g_2^3-27 g_3^2=0$,
this solution becomes a rational function of one exponential,
obtained by the degeneracy formula
\begin{eqnarray}
& &
\forall x,d:\
\wp(x,3 d^2,-d^3)
 = - d        + \frac{3 d}{2} \coth^  2  \sqrt{\frac{3 d}{2}} x.  
\label{eqwpcoth}
\end{eqnarray}
This codimension-two genus-zero solution is
\begin{eqnarray}
& & {\hskip -6.0 truemm}
\left\lbrace
\begin{array}{ll}
\displaystyle{
u=s_1 + b c + c \left(k^2 \tanh^2 k z - \frac{2}{3} k^2\right),\  
}\\ \displaystyle{
k^4=15 (s_1^2-s_2-(b c)^2)/c^2,\
}\\ \displaystyle{
s_3=\frac{1}{135} (3 s_1+3 b c+c k^2)
 (45 s_1^2-15 s_1 c (3 b+k^2)-4 c^2 k^4+30 b c^2 k^2-90 b^2 c^2),\
}\\ \displaystyle{
K=-\frac{2}{9} c^2 (3 b+k^2)^2 (45 b^2-30 b k^2+k^4),
}
\end{array}
\right.
\label{RSHsol1fam_trigo}
\end{eqnarray}
in which the arbitrary origin of $z$ has been omitted.

\subsection{Subequations obeyed by two Laurent series}
\label{subsectionTwoSeries}

The subequation then has degree $m=4$. 
One thus finds two subequations and only two.
One of them, as expected, is factorizable into the product of the two
second degree subequations (\ref{RSHP_subeq_1fam}) 
(one for $c$, the other for $-c$)
with the additional condition that the $K$ in (\ref{RSHP_subeq_1fam})
be invariant under parity on $c$, i.e.~$2 s_1^3 -3 s_2 s_1+ s_3=0$.

The other subequation is irreducible and has genus one,
\def\sstwo{\sigma_2}  
\def\ed{\sigma}
\begin{eqnarray}
& & {\hskip -12.0 truemm}
\left\lbrace
\begin{array}{ll}
\displaystyle{
F_4 \equiv c^2 \left({u'}^2+12 b ((u-s_1)^2 + 2 \sstwo)\right)^2
           -16 \left(             (u-s_1)^2 + 2 \sstwo)\right)^3=0,
}\\ \displaystyle{
\sstwo := \frac{1}{3} (5 (s_2-s_1^2) + 8 b^2 c^2),\
2 s_1^3 -3 s_2 s_1+ s_3=0,\
}\\ \displaystyle{
c^2 K=-\frac{16}{3} \left( 8 b^2 c^2 + 5 (s_2- s_1^2)\right)^2
                    \left(11 b^2 c^2 -10 (s_2- s_1^2)\right).
}
\end{array}
\right.
\label{RSHsubeq2fam}
\end{eqnarray}
The condition on $(s_1,s_2,s_3)$ expresses that the three fixed points
of (\ref{ODE_RCSH4}) are equispaced, like $-1,0,1$ in (\ref{ODE_RCSH0}).

In order to integrate this genus one equation $F_4(u,u')=0$, 
one must first establish a birational transformation
$(u,u') \leftrightarrow (\wp,\wp')$
between $F_4=0$ and the canonical equation of Weierstrass
$\wp'^2=4\wp^2-g_2\wp-g_3$,
i.e.~two pairs of rational transformations 
\begin{eqnarray}
u=R_1(\wp, \wp'), \qquad u'=R_2(\wp, \wp'), \qquad
\wp=R_3(u,u'), \qquad \wp'=R_4(u,u'),
\end{eqnarray}
where $R_i$, $i=1,2,3,4$ are rational in their arguments.
The algorithm consists of mapping (\textit{via} a birational transformation)
the given algebraic curve in $(u,u')$
to an algebraic curve in $(v,v')$ in which $v$ has a lower elliptic order than $u$,
until elliptic order $2$ (that of $\wp$) has been reached.
Outlined by Briot and Bouquet \cite[\S 250--251 p.~395]{BB}
for the so-called ``binomial'' and ``trinomial'' equations,
it has been implemented in the Maple package \verb+algcurves+ \cite{vanHoeij}
for any genus one first order equation.
Then a final scaling yields the desired solution.

This codimension-two elliptic solution of (\ref{ODE_RCSH3})
can be written in terms of either an even or an odd elliptic function
(just like $\sin(z-z_0)=-\cos(z-z_1),\ z_1-z_0=\pi/2$),
\begin{eqnarray}
& & {\hskip -6.0 truemm}
\left\lbrace
\begin{array}{ll}
\displaystyle{
u=s_1 -c (e_2-e_3)\frac{(\wp_0+b)^2 +g_2/4 -3 b^2}{\wp_0^2-b \wp_0-g_2/4+b^2}
 =s_1 - \frac{\D}{\D z} \left(\frac{c_1 c}{\wp_1-e_0}\right),
}\\ \displaystyle{
}\\ \displaystyle{
s_3 \hbox{ and } K \hbox{ as in }(\ref{RSHsubeq2fam}),
}\\ \displaystyle{
g_2= 2 (e_1^2 + e_2^2 + e_3^2)= 10 (s_1^2 - s_2 + 2 b^2 c^2)/(3 c^2),\ 
}\\ \displaystyle{
g_3= 4 e_1 e_2 e_3 =2 ( 5 (s_1^2-s_2) + 4 b^2 c^2) b /(3 c^2),\ 
}\\ \displaystyle{
(e_2-e_3)^2= \left(11 b^2 c^2 + 10 (s_1^2-s_2) \right),\ 
}\\ \displaystyle{
c_1^2=4 (e_0+b)^2 (2 e_0-b),\
\sstwo=2 c^2 (e_0+b)^2,
}\\ \displaystyle{
\Delta=g_2^3-27 g_3^2=\frac{4}{27 c^6}
      \left(11 b^2 c^2 + 10 (s_1^2-s_2) \right)
      \left( 8 b^2 c^2 -  5 (s_1^2-s_2) \right)^2,
}
\end{array}
\right.
\label{RSHsol2fam_elliptic}
\end{eqnarray}
in which $\wp_j$ is short for $\wp(z-z_j,g_2,g_3), j=0,1$,
and $z_0,z_1$ are constants of integration.
The equivalence between the conditions $K=0$ and $\Delta=0$ 
is a consequence of the suitable definition of $K$ in (\ref{ODE_RCSH3}).

The degeneracy $\Delta=0$ splits into two cases.
For $s_2=s_1^2 - (8/5)  (b c)^2$, the curve $F_4=0$ is reducible,
this is a particular case of subequation $F_2=0$.
For $s_2=s_1^2 +(11/10) (b c)^2$, the curve $F_4=0$ has genus zero,
therefore its solution is a rational function of one exponential,
readily obtained from the odd expression of $u$ in (\ref{RSHsol2fam_elliptic})
by the degeneracy formula (\ref{eqwpcoth}),
the final result 
is a solution of (\ref{ODE_RCSH3}) rational in hyperbolic tangent,
\begin{eqnarray}
& & {\hskip -6.0 truemm}
\left\lbrace
\begin{array}{ll}
\displaystyle{
u=s_1-\frac{2 c c_1  }{3 b}\frac{\D}{\D z} (1+\tanh^2 k z)^{-1} 
 =s_1-\frac{4 c c_1 k}{3 b}\frac{(1-\tanh^2 k z) \tanh k z}{(1+\tanh^2 k z)^2},\ 
}\\ \displaystyle{
k^2=\frac{3 b}{2},\
c_1^2=-54 b^3, 
}\\ \displaystyle{
K=0,\
s_2=s_1^2+\frac{11}{10} (b c)^2,\
s_3=s_1 \left(s_1^2 +\frac{33}{10} (b c)^2 \right), 
}
\end{array}
\right.
\label{RSHsol2fam_trigo}
\end{eqnarray}
in which the arbitrary origin of $z$ has been omitted.

The solutions 
(\ref{RSHsol1fam_elliptic}),
(\ref{RSHsol1fam_trigo}), 
(\ref{RSHsol2fam_elliptic}) and
(\ref{RSHsol2fam_trigo})
are thus all the meromorphic solutions of (\ref{ODE_RCSH3}),
and the last two ones, to the best of our knowledge, seem to be new\footnote{After submission, 
we were informed of similar results on (\ref{ODE_RCSH4}) by Kudryashov and Sinelshchikov \cite{KS2012} 
in which these authors apply another method.}.
These exact solutions, in particular the new ones, 
can be of an important practical use to check the validity of numerical simulations.

\section{Comparison with known results}
\label{sectionComparison}

Let us now examine how the complex solutions of (\ref{ODE_RCSH4})
found in section \ref{sectionMeromorphic_solutions} 
compare with some existing results 
on the real-valued smooth solutions of (\ref{ODE_RCSH0}). 
First we summarize the results obtained so far on (\ref{ODE_RCSH0})
for real $Z,U,a$. 

When $a>0$, 
\begin{itemize}
 \item \cite{Smets_vdB2002} 
       for almost all $a \in [0,2 \sqrt 2]$, 
       Eq.~(\ref{ODE_RCSH0}) admits a homoclinic solution in $H^2(\mathbb{R})$;
 \item \cite{SW2009} for all $a \in [0,a_0]$, where $a_0=1.228\dots$
 (see footnote on page \pageref{pagefootnote}).
 
       Eq.~(\ref{ODE_RCSH0}) admits a homoclinic solution $U$ such that $U-1\in H^2(\mathbb{R})$.
\end{itemize}

When $a\leq0$, any bounded smooth solution of (\ref{ODE_RCSH0}) 
has the following properties \cite{vandenBerg1998}: 
\begin{itemize}
 \item $a\leq 0\Rightarrow |U|\leq \sqrt{2}$;
 \item $a\leq -2 \sqrt{2} \Rightarrow |U|\leq 1$.
\end{itemize}

Let us reduce (\ref{ODE_RCSH4}) to (\ref{ODE_RCSH0}) 
by assuming both independent variables 
$Z$ and $z$ to be equal and real ($\mu=1$ in (\ref{scalingtransfo})).  
Let us restrict the domain of the four solutions of (\ref{ODE_RCSH0})
associated to
(\ref{RSHsol1fam_elliptic}),
(\ref{RSHsol1fam_trigo}), 
(\ref{RSHsol2fam_elliptic}) and
(\ref{RSHsol2fam_trigo}) 
to the real line and examine whether those resulting solutions 
possess one or more of the following properties:
real-valuedness, boundedness, homoclinic topology. 

Let us first examine the two genus-one solutions of (\ref{ODE_RCSH0}).
Easily deduced from 
(\ref{RSHsol1fam_elliptic}) and 
(\ref{RSHsol2fam_elliptic}), they are respectively
\begin{eqnarray}
& & {\hskip -6.0 truemm}
U=\pm \sqrt{-120} \left(\wp(Z,g_2,g_3)+\frac{a}{60}\right),\
g_2=-\frac{a^2+10}{180},\
g_3=\frac{a(a^2+5)}{5400},
\label{RSHsol1fam_ellipticU}
\end{eqnarray}
and 
\begin{eqnarray}
& & {\hskip -6.0 truemm}
\left\lbrace
\begin{array}{ll}
\displaystyle{
U=\pm \sqrt{\frac{100 - 11 a^2}{90}} 
   \frac{(a^2+25-360a \wp(Z,g_2,g_3)-10800 \wp(Z,g_2,g_3)^2)}
       {2 a^2-25+180a \wp(Z,g_2,g_3)-10800 \wp(Z,g_2,g_3)^2},
}\\ \displaystyle{
}\\ \displaystyle{
g_2=\frac{a^2-5}{540},\
g_3=\frac{2 a^3-25 a}{162000},
}\\ \displaystyle{
}\\ \displaystyle{
e_1=-\frac{a}{60},\
e_2=\frac{3 a+\sqrt{3(11 a^2 -100)}}{360},\
e_3=\frac{3 a-\sqrt{3(11 a^2 -100)}}{360},\
}
\end{array}
\right.
\label{RSHsol2fam_ellipticU}
\end{eqnarray}
where the arbitrary origins of $Z$ have been omitted. 
For the above two solutions,
if $a$ is real, then $g_2$ and $g_3$ are also real.
By Theorem 3.16.2 in \cite{JS1987}, 
$\wp$ (and therefore $\wp'$) 
is then real-valued for all $Z\in \mathbb{R}$ and $\wp$ has a real period. 
This implies that the elliptic solution
(\ref{RSHsol1fam_ellipticU}) is not real-valued
and that the solution (\ref{RSHsol2fam_ellipticU})
is real-valued if and only if $a^2 < 100/11$. 

Let us now prove that, when it is real, the solution (\ref{RSHsol2fam_ellipticU})
is bounded on $\mathbb{R}$.
The necessary conditions for $U$ to be extremal are
either $\wp(Z)=\infty$ or $\wp'(Z)=0$
(which implies $\wp(Z)=-a/60$ since $e_2,e_3$ are not real 
and ${\wp'}^2=4(\wp-e_1)(\wp-e_2)(\wp-e_3)$) or
\begin{eqnarray}
& & {\hskip -6.0 truemm}
270 a \wp^2(Z)+(a^2-50) \wp(Z) +\frac{a(a^2-5)}{24}=0.
\label{RSHsol2fam_elliptic_extrema1}
\end{eqnarray}

\begin{figure}[ht]
\begin{center}
\includegraphics[width=0.3\linewidth,height=0.2\textheight]{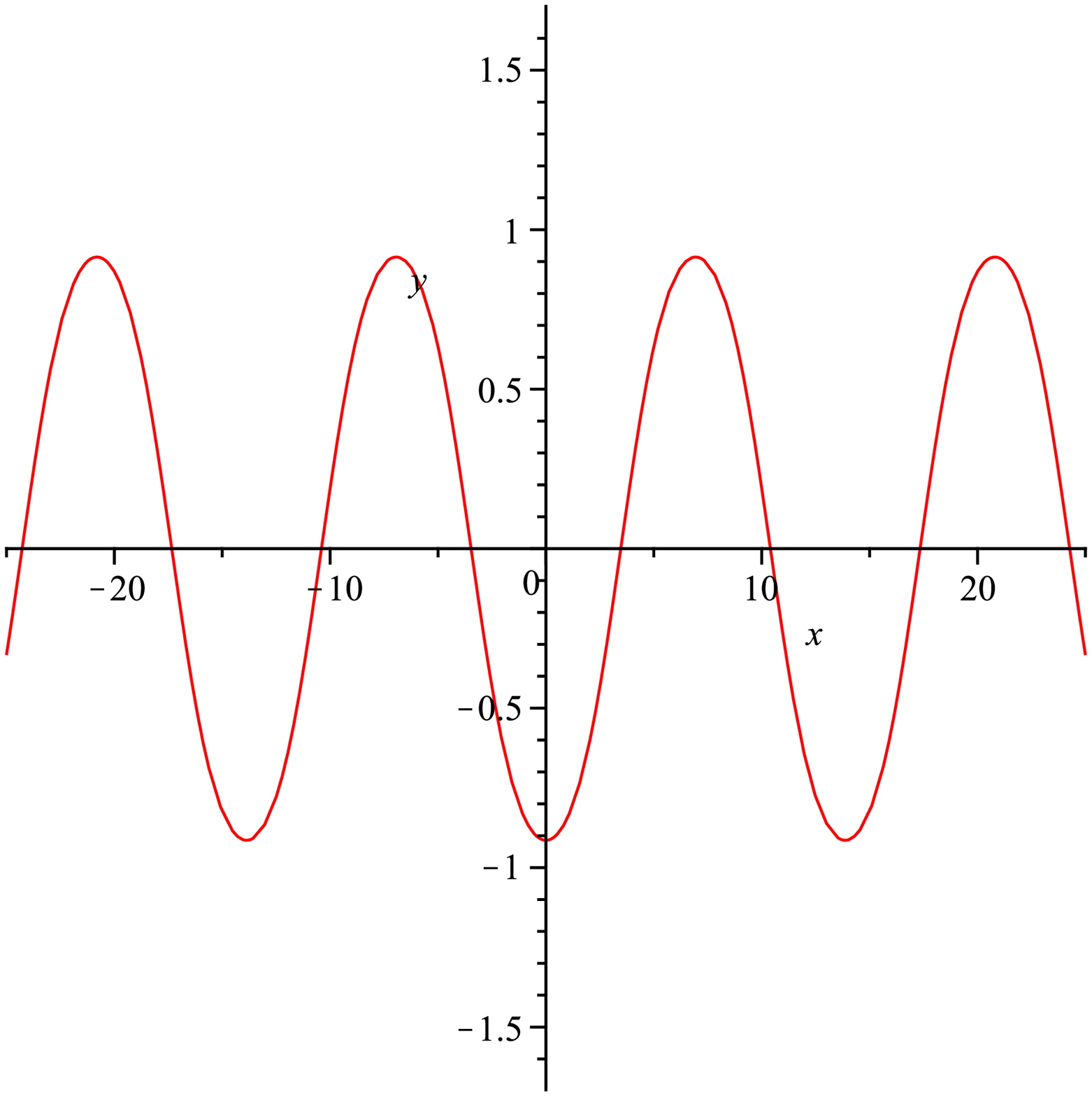}
\includegraphics[width=0.3\linewidth,height=0.2\textheight]{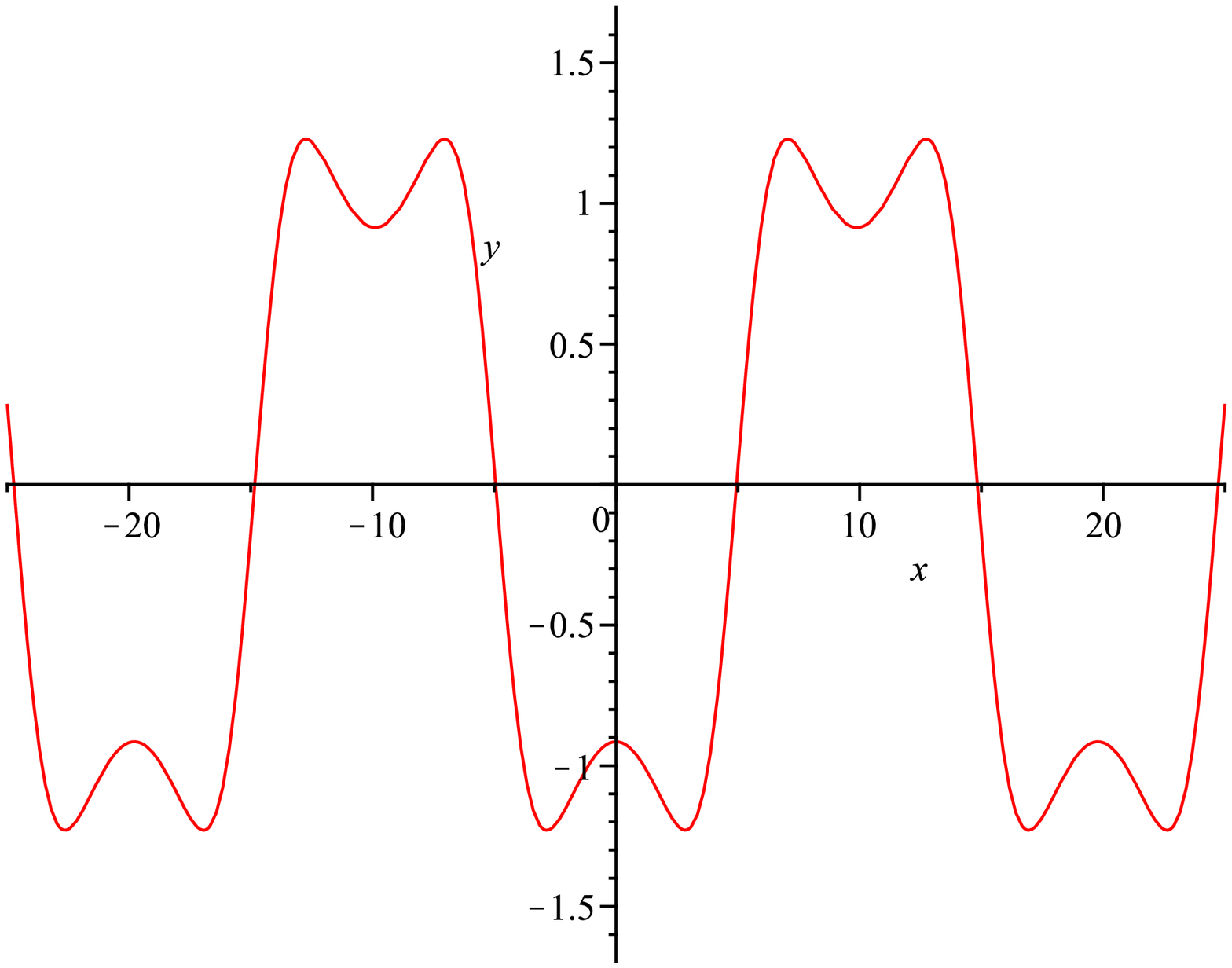}
\caption{ 
\hfill\break \footnotesize Elliptic solution $U(Z)$ of (\ref{ODE_RCSH0}) defined by 
(\ref{RSHsol2fam_ellipticU}), from left to right:
\hfill\break  (i) $a= -1.5, b=a/60, g_2\approx 0.0134, g_3\approx 2.731\times 10^{-4},
\Delta\approx 3.6499\times 10^{-7}$;
\hfill\break (ii) $a=  1.5, b=a/60, g_2\approx 0.0134, g_3\approx 2.731\times 10^{-4}, 
\Delta\approx 3.6499\times 10^{-7}$.
}
\label{Fig_elliptic} 
\end{center}
\end{figure}

The last event (\ref{RSHsol2fam_elliptic_extrema1}) can only occur for $0<a<\sqrt{100/11}$. 
It is because from ${\wp'}^2=4(\wp-e_1)(\wp-e_2)(\wp-e_3)$ and $\wp-e_2=\overline{(\wp-e_3)}$, 
we know that $\wp -e_1 \ge 0$ as $\wp'$ is real on $\mathbb{R}$. 
Therefore $\wp$ attains the global minimun value $e_1=-a/60$ when it is restricted on the real line. 
Now if $-\sqrt{100/11} < a \le 0$, then it can be checked easily 
that the two zeros of the equation $270 a p^2+(a^2-50) p +\frac{a(a^2-5)}{24}=0$ 
are both less than $e_1$ 
and this implies that $\wp$ cannot satisfy (\ref{RSHsol2fam_elliptic_extrema1}) 
on the real line if $-\sqrt{100/11} < a \le 0$.

\begin{eqnarray}
\left\lbrace
\begin{array}{ll}
\displaystyle{
 U_{1,-}=\mp \sqrt{\frac{10}{9}-\frac{11 a^2}{90}} \quad \text{when} \quad \wp(Z)=-\frac{a}{60},
}\\ \displaystyle{
 U_{1,+}=\pm \sqrt{\frac{10}{9}-\frac{11 a^2}{90}} \quad \text{when} \quad \wp(Z)= \infty,
}\\ \displaystyle{
 U_{2,\pm}=\pm \sqrt{\frac{10}{9}+\frac{8 a^2}{45}}
\quad \text{when} \quad \wp(Z)=\frac{50 -a^2 \pm \sqrt{(100-11 a^2)(4 a^2+25)}}{540 a}.
}
\end{array}
\label{RSHsol2fam_elliptic_bound3}
\right.
\end{eqnarray}
Therefore, whenever it is real-valued,
the solution (\ref{RSHsol2fam_ellipticU}) is bounded on $\mathbb{R}$,
with the following bounds for $U^2$ (because of parity, it is sufficient to consider $U^2$),
\begin{eqnarray}
\left\lbrace
\begin{array}{ll}
\displaystyle{
-10/ \sqrt{11} < a \le 0:\ 0 \le U^2 \le \frac{10}{9}-\frac{11 a^2}{90},
}\\ \displaystyle{
0 \le a <  +10/ \sqrt{11}:\ 0 \le U^2 \le \frac{10}{9}+\frac{8 a^2}{45}.
}
\end{array}
\label{RSHsol2fam_elliptic_bound4}
\right.
\end{eqnarray}
Within one real period, the graph of $U(Z)$ (see Figure \ref{Fig_elliptic})
displays either two opposite extrema (case $-10/ \sqrt{11} < a \le 0$)
or six extrema (case $0 \le a < 10/ \sqrt{11}$)
made of two extrema $U_{1,\pm}$ and four extrema $U_{2,\pm}$.

All these results agree with those in \cite{vandenBerg1998}
mentioned at the beginning of this section. 

Let us now examine whether the solutions 
(\ref{RSHsol1fam_trigo}) and (\ref{RSHsol2fam_trigo}) of (\ref{ODE_RCSH0}) 
are real and can represent a bounded or homoclinic solution. 
The first genus zero solution (\ref{RSHsol1fam_trigo})
when converted to a solution $U(Z)$ of (\ref{ODE_RCSH0}) 
implies a constraint on $a$, namely, 
\begin{eqnarray}
& & 
(4 a^2+25)(8 a^4+ 55 a^2 + 200)=0,
\end{eqnarray}
which admits no real solution for $a$, so this solution must be discarded.

The meromorphic solution $U(Z)$ of (\ref{ODE_RCSH0})
defined by the second genus zero solution (\ref{RSHsol2fam_trigo}) is,
\begin{eqnarray}
& & {\hskip -6.0 truemm}
U=\pm \frac{4 \sqrt{30}}{\sqrt{11}} 
  \frac{(1-\tanh^2 k Z) \tanh k Z}{(1+\tanh^2 k Z)^2},\
k^4=\frac{1}{176},\ a=40 k^2, 
\label{RSHsol2fam_trigoU}
\end{eqnarray}
where the arbitrary origin has been omitted.
When restricted to the real $Z$ line, 
this solution is real-valued if and only if $k^2=1/(4 \sqrt{11})$ and $a=10/ \sqrt{11}$.
It is then bounded on $\mathbb{R}$ and homoclinic,
and displays one minimum and one maximum,
$U_{\rm extrema}=\pm \sqrt{30/11}\approx 1.65 $,
reached for $\tanh k Z=\varepsilon (\sqrt{2}-1)$, where $\varepsilon^2=1$,
in agreement with the limit $a \to 10/ \sqrt{11}$ in (\ref{RSHsol2fam_elliptic_bound3}).

\begin{figure}[ht]
\begin{center}
\includegraphics[width=0.3\linewidth, height=0.2\textheight]{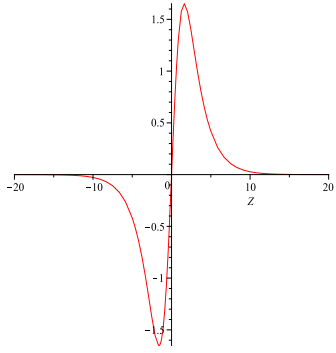} 
\caption{ 
\hfill\break \footnotesize Trigonometric solution U(Z) of (\ref{ODE_RCSH0}) defined by 
(\ref{RSHsol2fam_trigoU}) when $a=10/\sqrt{11}, k=1/(2\times 11^{1/4})$. \,
}
\label{Fig_trigo} 
\end{center}
\end{figure}

To summarize, the exact solution 
(\ref{RSHsol2fam_trigoU}) with $k^2=1/(4 \sqrt{11})$ and $a=10/ \sqrt{11}$ 
is the unique meromorphic homoclinic solution 
(up to a translation of origin) 
mentioned in the introduction. 
See Figure \ref{Fig_trigo}.

The bound of the solution (\ref{RSHsol2fam_ellipticU}) is an 
increasing function of $a$.
It would be interesting to know whether the same trend will be observed for
the class of all bounded smooth solutions of (\ref{ODE_RCSH0}),
and whether the bound $\sqrt{30/11}$ is sharp 
for any bounded smooth solution whenever $0<a<10 / \sqrt{11}$.

\remark
We could not obtain a codimension zero singlevalued solution of (\ref{ODE_RCSH3}),
If it exists, this solution,
which is \textbf{not} meromorphic and therefore not elliptic, 
is locally represented by the two Laurent series (\ref{SHlau}),
which excludes by construction the contribution of the irrational
Fuchs indices $(7 \pm \sqrt{-71})/2$.
Such series have been shown to be convergent by Chazy \cite{ChazyThese}.
Although there is no analytic evidence of the existence of this nonmeromorphic solution,
a numerical study by Pad\'e approximants in the similar situation 
of the traveling wave reduction of the Kuramoto-Sivashinsky equation
\cite{YCM2003,CMBook} does not yield any evidence of its nonexistence either.
It could be possible that this singlevalued, nonmeromorphic solution
displays a movable natural boundary.
\end{preremark}

\end{document}